\documentclass[reprint,nofootinbib,nobibnotes,amsmath,amssymb,aps,prb,floatfix]{revtex4-2}
\fontfamily{cmss}\selectfont 
\usepackage{graphicx}
\usepackage{dcolumn}
\usepackage{bm}
\usepackage{tabularx}
\usepackage{xcolor}
\usepackage{overpic}
\usepackage{multirow}
\usepackage{hyperref}
\usepackage{braket}

\newcommand{\beq}{\begin{eqnarray}}
\newcommand{\eeq}{\end{eqnarray}}

\newcommand{\figdisp}[1]{Fig. \ref{#1}}

\newcommand{\ignore}[1]{}

\begin{document}

\preprint{APS/123-QED}

\title{Solving the SU(\texorpdfstring{$N$}{}) Orbital Hatsugai-Kohmoto Model}

\author{Nico A. Hackner}
\author{Peizhi Mai}
\author{Philip W. Phillips}
 \affiliation{Department of Physics and Anthony J. Leggett Institute for Condensed Matter Theory, University of Illinois Urbana-Champaign, Urbana, Illinois 61801, USA}

\begin{abstract}
We show that the physics of the SU($N$) Hubbard model can be realistically simulated with the recently developed orbital Hatsugai-Kohmoto model.  In this approach, the momentum mixing absent from the band Hatsugai-Kohmoto model is included by grouping $n$-Hubbard atoms into a cluster.  We take advantage of the rapid convergence of this scheme ($1/n^2$) and show that for $n\approx O(10)$, quantitative agreement with the determinantal
quantum Monte Carlo method arises for the double occupancy as well as qualitative agreement for the filling and compressibility across the Mott transition. Additional features of this work are that we can obtain low-temperature physics, which could be qualitatively different from its high-temperature counterpart, and dynamical quantities without resorting to analytical continuation, thereby establishing the orbital Hatsugai-Kohmoto model as a useful alternative perspective for studying strong correlations in SU($N$) systems.
\end{abstract}

-\maketitle
 

\section{\label{sec:intro}Introduction}
While SU($2$) is the standard group relevant for describing electronic matter, ultra-cold alkaline earth atoms~\cite{GorshkovNphys2010,CazalillaRPP2014} can realize SU($N$) symmetry by having nuclear spins as large as 10 and in some cases 36 in Na$^{40}$K~\cite{SchindewolfNature2022}. Consequently, the more general group SU($N$)~\cite{honerkamp2004,SchindewolfNature2022,wu_2006,cazalilla_2009,GorshkovNphys2010,taie_2012,CazalillaRPP2014,hofrichter2016,ozawa_2018,wang_slater_2019,taie2022,tusi2022,ibarra_2024,Pasqualetti2024} is highly relevant to the cold-atom community. Here SU($N$) acts globally transforming any two local eigenstates $|\alpha,i\rangle$  and $|\beta,i\rangle$ via the unitary matrix $U_{\rm\alpha,\beta}$, such that 
$|\alpha,i\rangle=\sum_\beta U_{\alpha\beta}|\beta, i\rangle$.  If the interactions are the same for all $N$ spin components, realizing flavor-selective Mott physics is possible regardless of the underlying lattice.  Such a realization obviates the complications introduced by multiple orbitals in traditional solid-state systems that exhibit orbital-selective Mott physics~\cite{MediciPRL2014,YiNC2015}. In terms of magnetism, SU($N$) Hubbard and Heisenberg models~\cite{Hermele2009, toth2010,Bauer2012, Hermele2011, romen2020,Nataf2014,Sotnikov2014,Sotnikov2015,Hafez2018,Hafez2019,feng2023,ibarra2023} provide a richer playground than does SU($2$).  While a spin pattern with a period of $N$ is the most generic ordering in SU($N$), stripe phases also obtain, for example, a dimerized pattern not formed out of singlets of SU($4$) but rather a six-dimensional irreducible representation in SU($4$)~\cite{corboz}.  That is, the stripe pattern is not composed of ladders of dimers forming singlets, but rather a set of gapless excitations emerge from the breaking of the SU($4$) despite the accompanying reduction in the translational symmetry~\cite{corboz}.  SU($N$) physics is not limited to single atomic nuclei and extends, for example, to molecular systems as well.  Recent experiments~\cite{SchindewolfNature2022} demonstrate that because $s$-wave collisions of shielded molecules have a limited dependence on the underlying spin states, such systems provide a perfect playground for exploring  SU($N$) physics.  Although the interaction between such molecules is of the standard dipolar form, the sign and magnitude of the scattering length can be tuned by the details of the shielding field.  As a result, shielded molecular aggregates from atomic nuclei can even exhibit bosonic or fermionic statistics.

Coupled with the tunability of the interactions, simply viewing the spin states as color degrees of freedom makes it possible to study analogues of quark confinement in shielded ultra-cold atoms.   To see this, consider the dipole-dipole interaction
\beq
\hat{H}_{dd}=\frac{-3({\bf \mu}_1\cdot\hat{\bf R})({\bf \mu}_2\cdot\hat{\bf R})-{\bf \mu}_1\cdot{\bf \mu}_2}{4\pi\epsilon_0 R^3}, 
\eeq
between two dipolar molecules of magnitude $\mu_1$ and $\mu_2$ separated by a distance $R$.  For $s$-wave scattering, although this interaction averages to zero for total angular momentum $L=0$, matrix elements connect the $L=0$ and $L=2$ states that have both diagonal and off-diagonal components.  The off-diagonal component is attractive and scales~\cite{anisot} as $d^4/R^4$ where $d$ is the space-fixed dipole moment of each molecule. The repulsive interaction arises specifically from the energy shift, $\Delta E$, induced by the external static field and microwave shielding fields.  The matrix element connecting pair states is proportional to $d^4/(\Delta E R^6)$ and hence is shorter range relative to the attractive interaction. It is the interplay between the short-range repulsion and long-range attraction that generates confined and deconfined phases that are color specific~\cite{hazard2}.

While the knee-jerk instinct is that SU($N$) in the large $N$ limit is inherently mean-field, this is not the case for describing the details of the phase diagram for ultra-cold atoms.  Indeed, discerning the phase diagram of SU($N$) matter is particularly taxing computationally, as the size of the Hilbert space scales as $2^{N\times N_{\rm sites}}$.  Consequently, advancing the physics of SU($N$) systems requires new tools.  In this study, we adopt the recent orbital Hatsugai-Kohmoto (OHK) model~\cite{dmitry2023, mai2024}.  This model builds in momentum mixing into the tractable HK model~\cite{HKJPSJ1992,HKnp1,HKnp2} (now referred to as band HK or $1$-OHK). If $n$ is the number of momenta that are mixed, the convergence to Hubbard scales as $1/n^2$.  Such rapid convergence makes this a tractable scheme for simulating Hubbard physics. It is this scheme that we employ here for the SU($N$) problem. We compute the double occupancy, compressibility, Mott gap, and dynamical spin structure factor across the Mott transition. We obtain the latter directly from exact diagonalization without the need for analytical continuation as is required in quantum Monte Carlo methods. Our results show quantitative agreement with these numerical approaches where comparison is possible, and we extend our analysis to regions where these methods become impractical.

\begin{figure}
    \begin{overpic}[width=0.9\columnwidth, keepaspectratio]{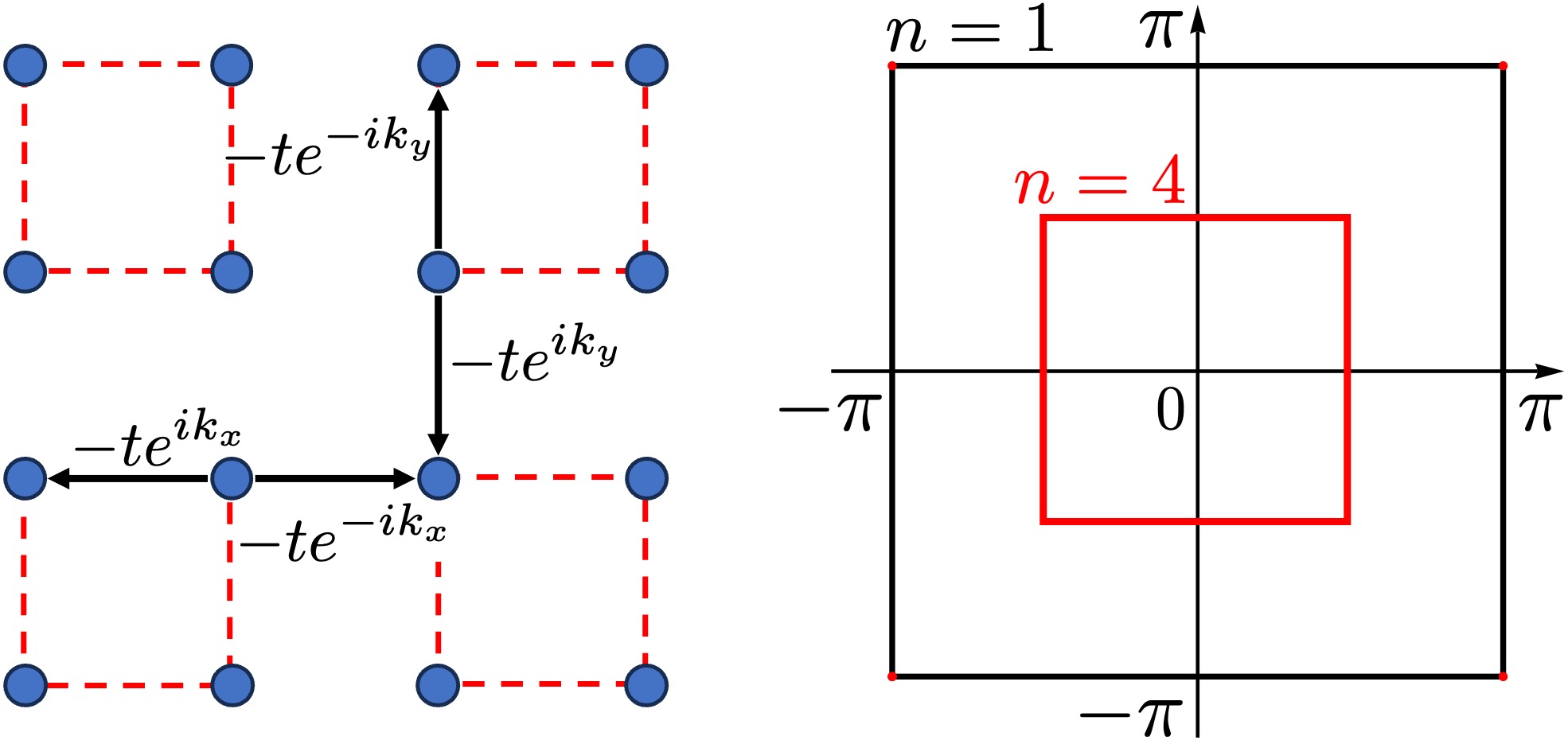}
    \put(-8,  44){(a)}
    \put(48,  44){(b)}
  \end{overpic}
    \caption{Panel (a) is a schematic showing the real-space clustering for 4-OHK and some representative elements of the hopping matrix $\xi^{\alpha\alpha'}_\mathbf{k}$. Panel (b) shows the original BZ, i.e. the $n=1$ limit, and the rBZ$_4$ for 4-OHK (for a $2\times 2$ unit cell). }
	\label{fig:clustering}
\end{figure}

\begin{figure*}[ht]
    \begin{overpic}[width=0.9\textwidth, keepaspectratio]{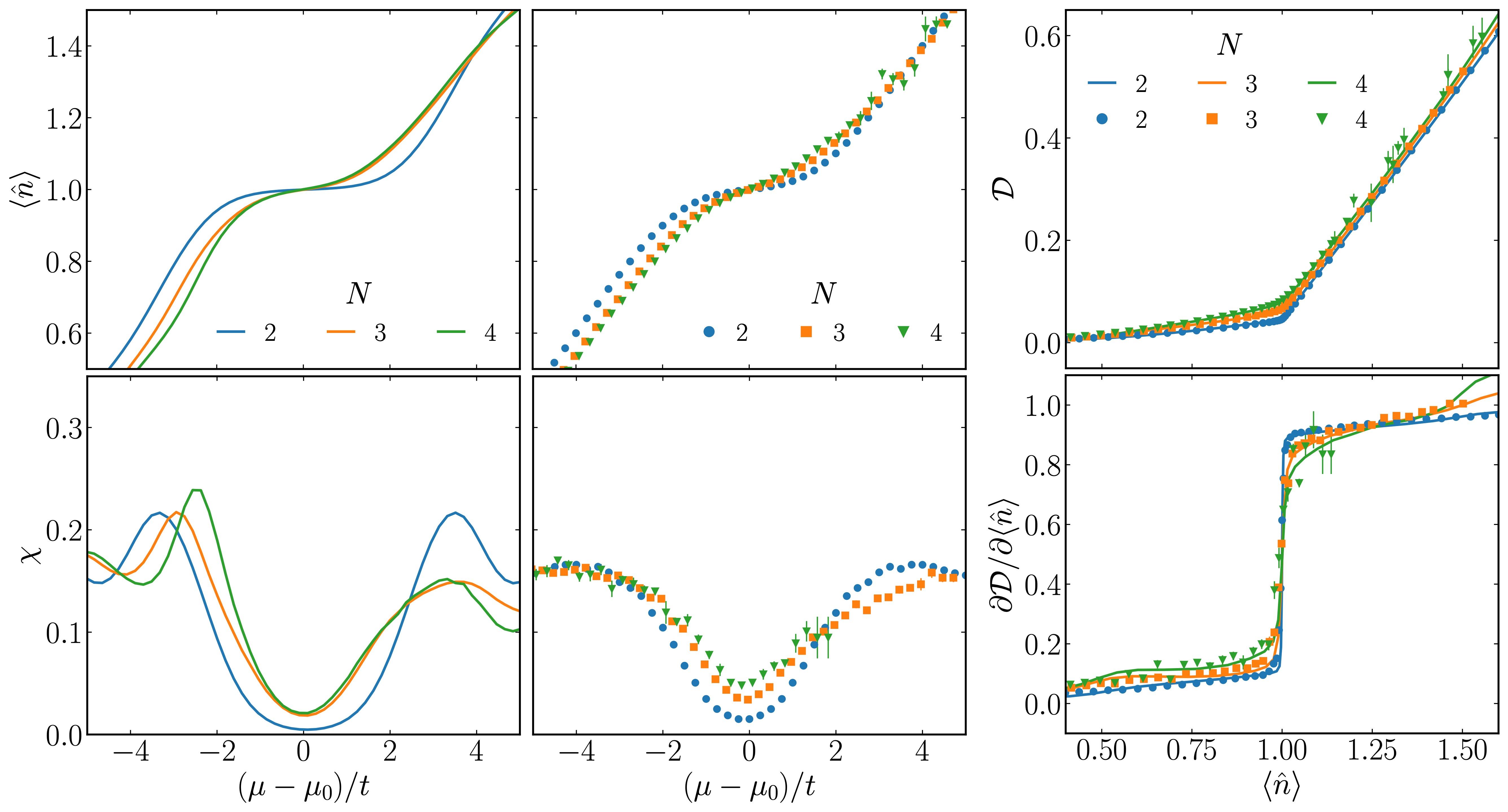}
    \put(17, 54.5){4-OHK}
    \put(46.5, 54.5){DQMC}
    \put(7, 50.5){(a)}
    \put(36.5, 50.5){(b)}
    \put(7, 26.5){(c)}
    \put(36.5, 26.5){(d)}
    \put(72, 50.5){(e)}
    \put(72, 26.5){(f)}
  \end{overpic}
	\caption{Panels (a-d) show density $\langle \hat{n} \rangle$ and compressibility $\chi= \partial \langle \hat{n} \rangle / \partial \mu$ versus chemical potential $\mu$ with SU($N$) 4-OHK in panels (a,c) and SU($N$) Hubbard obtained with DQMC~\cite{ibarra2021} in panels (b,d). $\mu_0$ is calculated such that $\langle \hat{n} \rangle |_{\mu_0} = 1$. We fix $U=8t$ and $\beta = 2t^{-1}$, and find $\mu_0/U =0.49,0.40,0.35$ for $N=2,3,4$, respectively. Panels (e) and (f) show double occupancy $\mathcal{D}$ and its derivative $\partial \mathcal{D} / \partial \langle \hat{n} \rangle$ versus $\langle \hat{n} \rangle$, respectively. Here we fix $U=8t,\ \beta = 2t^{-1}$. The line and scatter plots represent 4-OHK and DQMC-Hubbard results~\cite{ibarra2021}, respectively.}
	\label{fig:density_compress}
\end{figure*}

\section{\label{sec:model}Model and methods}

This work aims to establish OHK models as an efficient tool for simulating SU($N$) Hubbard physics. We begin with a review of the connection between the Hubbard model and OHK models following the discussion in Ref.~\cite{mai2024}. Consider first the Hubbard interaction
\begin{align}
    H_\text{HB}^\text{int} &= U \sum_i n_{\mathbf{r}_i\uparrow} n_{\mathbf{r}_i\downarrow} \nonumber\\
    &= \frac{U}{N_s}\sum_{\mathbf{k,p,q}\in \text{BZ}} c^\dagger_{{\bf k}\uparrow} c_{{\bf k}-\bf q\uparrow}c^\dagger_{{\bf k}+{\bf p}\downarrow}c_{{\bf k}+{\bf p}+{\bf q}\downarrow} ,
\end{align}
where $N_s$ is the number of sites, and the band HK interaction
\begin{align}\label{eq:band}
    H^\text{int}_\text{HK} = U \sum_\mathbf{k\in \text{BZ}}  c^\dagger_{\mathbf{k}\uparrow} c_{\mathbf{k} \uparrow} c^\dagger_{\mathbf{k} \downarrow} c_{\mathbf{k} \downarrow} .
\end{align}
 Note that we restrict our attention to SU(2) for simplicity. The momentum space representation of the Hubbard interaction highlights the deviation between the two models, namely, band HK contains no momentum scattering. We can move towards Hubbard physics by systematically introducing scattering momenta ${\bf p, q}\in B_n$ to the interaction term, leading to the OHK model. We refer to the OHK model with $n$ scattering momenta (or equivalently, $n$ orbitals, as illustrated below) as the $n$-OHK model. The case of $n=4$ is depicted in Fig.~\ref{fig:clustering}. Here we set $B_4 = \{ (0,0), (0,\pi), (\pi,0), (\pi,\pi) \}$ and consider an OHK interaction
\begin{align}\label{eq:hubbard-like}
    H^{\text{int}, n=4}_\text{OHK} = \frac{U}{4}\sum_{\mathbf{k}\in \text{BZ}} \sum_{\mathbf{p,q}\in B_4} c^\dagger_{{\bf k}\uparrow} c_{{\bf k}-\bf q\uparrow}c^\dagger_{{\bf k}+{\bf p}\downarrow}c_{{\bf k}+{\bf p}+{\bf q}\downarrow}. 
\end{align} 
The corresponding real-space picture is shown in Fig.~\ref{fig:clustering}(a), where each unit cell includes four sites/orbitals. We can consider a transformation to the orbital basis $c_{\alpha\mathbf{k}\sigma}=\frac{1}{2}\sum_{\mathbf{q}\in B_4}c_{\mathbf{k+q}\sigma}e^{-i\mathbf{q}\cdot\mathbf{r_\alpha}}$, where $\alpha=1,...,4$ labels the orbitals and $\bf r_\alpha$ denotes the position of each orbital within the unit cell. Performing this transformation on Eq.~\ref{eq:hubbard-like} leads to
\begin{align}
    H^{\text{int}, n=4}_\text{OHK} = U \sum_{\alpha=1}^4 \sum_{\mathbf{k}\in \text{rBZ}_4} c^\dagger_{\alpha \bf k \uparrow} c_{\alpha \bf k \uparrow} c^\dagger_{\alpha \bf k \downarrow} c_{\alpha \bf k \downarrow},
\end{align}
where the momentum summation is restricted to a reduced Brillouin Zone (rBZ$_n$), reflecting the enlarged unit cell in real space (see Fig.~\ref{fig:clustering}(b)). We see that the interaction term reduces to a form similar to Eq.~\ref{eq:band}, however, with additional orbital degrees of freedom. This completes the construction of the interaction term of the 4-OHK model. This construction can be generalized to an arbitrary number of scattering momenta, and Hubbard physics in the thermodynamic limit is guaranteed to be obtained as $n\rightarrow\infty$~\cite{mai2024}. The $n^2$ terms in Eq.~\ref{eq:hubbard-like} introduced by $n$ scattering momenta, along with the numerical evidence from \cite{mai2024}, indicate that the convergence to Hubbard scales at least as fast as $1/n^2$. Therefore, $n$ does not need to be excessively large for quantitative agreement with the Hubbard model. Note that while the addition of momentum scattering modifies the interaction, we ensure that the kinetic term in the Hamiltonian is kept unchanged throughout this process
\begin{align}
    \sum_{\bf k \in \text{BZ}, \sigma} \xi_{\bf k} c^\dagger_{\bf k \sigma} c_{\bf k \sigma} = \sum_{\substack{\mathbf{k}\in{\rm rBZ}_n,\\ \alpha\alpha',\sigma}} \xi^{\alpha\alpha'}_{\bf k} c^\dagger_{\alpha \bf k \sigma} c_{\alpha' \bf k \sigma}. 
\end{align}
 Here $\xi^{\alpha\alpha'}_{\bf k}=\sum_{\delta\mathbf{r}} t^{\alpha \alpha'}_{\delta \mathbf{r}} e^{i\mathbf{k}\cdot(\mathbf{\delta r + r_\alpha-r_{\alpha'}})}  - \mu \delta_{\alpha,\alpha'}$, where $t^{\alpha \alpha'}_{\delta \mathbf{r}}$ is the hopping between $\alpha$ and $\alpha'$ orbitals in unit cells separated by $\delta \mathbf{r}$, and $\mu$ is the chemical potential. In this work, we consider a two-dimensional square lattice with nearest-neighbour hopping, where $\xi_{\bf k} = -2 t (\cos{k_x} + \cos{k_y})-\mu$, with $t=1$ setting the energy scale. Representative off-diagonal elements of $\xi^{\alpha\alpha'}_{\bf k}$ for the 4-OHK model are shown schematically in Fig.~\ref{fig:clustering}(a).  Note that, as a result of the orbital mixing, the kinetic and potential terms no longer commute.  It is this new ingredient that introduces spectral weight transfer on the Mott scale.

The SU($N$) generalization\cite{nogueira1996}  of the band HK model is exactly solvable and was introduced soon after the original band HK model(see the appendix for a discussion of this model). The introduction of momentum mixing to the SU($N$) HK model follows straightforwardly from the SU(2) case, allowing us to introduce the SU($N$) OHK model
\begin{align}\label{eq:SU(N)OHK}
    H_\text{OHK}^N =& \sum_{\substack{{\bf k}\in{\rm rBZ}_n,\\ \alpha\alpha', \sigma}} \xi^{\alpha\alpha'}_{\bf k}  c^\dagger_{\alpha {\bf {\bf k} } \sigma}c_{\alpha' {\bf k} \sigma}\nonumber\\
    &+ \frac{U}{2} \sum_{\substack{{\bf k}\in{\rm rBZ}_n,\\\alpha, \sigma\neq\sigma'}} n_{\alpha {\bf k} \sigma} n_{\alpha {\bf k} \sigma'},
\end{align} 
where $\alpha,\alpha'=1,..., n$ denote orbital indices and $\sigma,\sigma'=1,...,N$ denote the flavor degree of freedom. In the case of SU($N$), the inclusion of $n$ orbitals introduces $N(N-1)n^2/2$ terms when expressing the interaction in the form of Eq.~\ref{eq:hubbard-like}. This suggests faster convergence to Hubbard physics with increasing $n$ when compared to the SU(2) case.

All numerical results presented in this work are obtained using exact diagonalization of the Hamiltonian in Eq.~\ref{eq:SU(N)OHK}. We obtain dynamical quantities by applying the Lanczos method to the ground state obtained from exact diagonalization. The 4-OHK calculations are simulated on the rBZ$_4$ with $L \times L$ points where $L=20$. For 8- and 9-OHK calculations, we set $L=10$.

\begin{figure*}
  \begin{overpic}[width=0.9\textwidth, keepaspectratio]{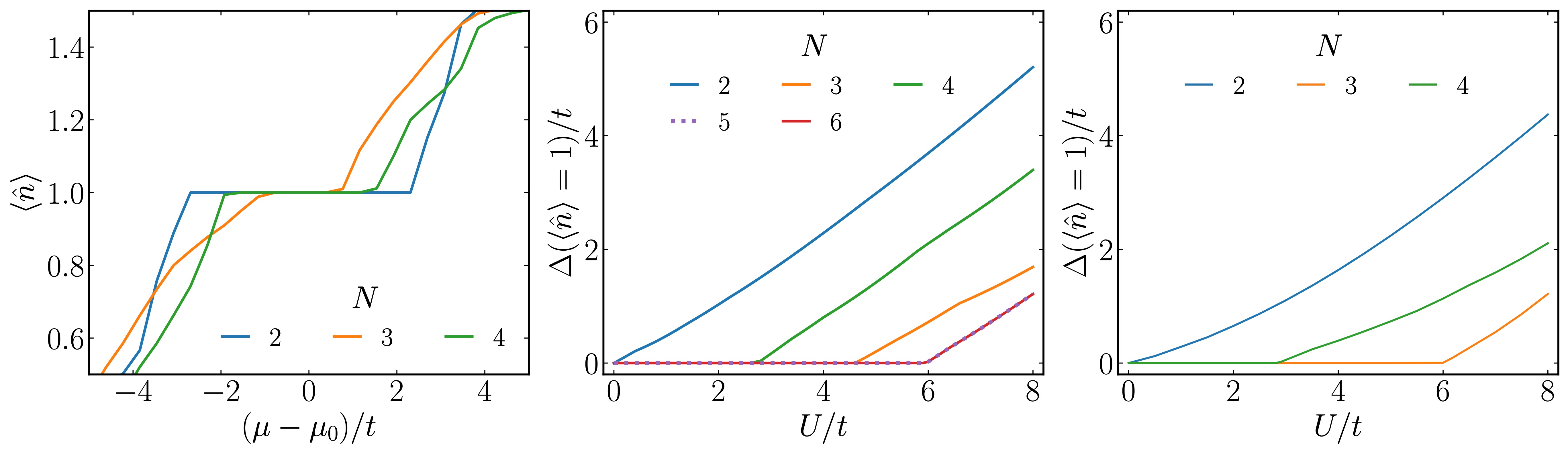}
    \put(6.5,  26){(a) 4-OHK}
    \put(39.2,  26){(b) 4-OHK}
    \put(72,  26){(c) 8-OHK}
  \end{overpic}
    \caption{Panel (a) shows the density $\langle \hat{n} \rangle$ versus chemical potential $\mu$ at $U=8t$ and $\beta = 200t^{-1}$ for varying $N=2,3,4$ in SU($N$) 4-OHK models. Panels (b) and (c) show the charge gap $\Delta$ at $\langle \hat{n} \rangle=1$ filling as a function of interaction strength $U$ at zero temperature for varying $N$ in SU($N$) 4- and 8-OHK models, respectively. For the 8-OHK model, we use a $2\times 4$ unit cell.}
	\label{fig:mott_gap}
\end{figure*}

\section{\label{sec:thermo} Results of SU($N$) OHK}

We begin by investigating the generalized SU($N$) OHK models with a focus on their general $N$-dependent behaviors. For a given $N$, Mott insulating states are possible at fillings $\langle \hat{n} \rangle = a$ for $a=1,...,N$. Of particular interest are the cases of one particle per site, $\langle \hat{n} \rangle = 1$, and half-filling, $\langle \hat{n} \rangle = N/2$ (when $N$ is even). These two phases coincide for $N=2$ and hence they are natural SU($N$) generalizations of the SU(2) half-filled Mott insulator.

\subsection{Mott insulator at \texorpdfstring{$\langle \hat{n} \rangle = 1$}{}}

{\bf High temperature.} We first solve 4-OHK (using a $2\times 2$ unit cell), at high temperature $\beta=2t^{-1}$ enabling a direct comparison with previous determinantal quantum Monte Carlo (DQMC) simulations on a $6\times6$ SU($N$) Hubbard cluster~\cite{ibarra2021}. In Fig.~\ref{fig:density_compress}, we show the filling
\begin{align}
    \langle \hat{n} \rangle = \frac{1}{n V} \sum_{\substack{\mathbf{k}\in{\rm rBZ}_4,\\ \alpha,\sigma}} \langle n_{\alpha \mathbf{k} \sigma} \rangle ,
\end{align}
and compressibility $\chi\equiv \partial \langle \hat{n} \rangle / \partial \mu$ as a function of $\mu$ for the 4-OHK model in panels (a) and (c), and the corresponding DQMC results are shown in panels (b) and (d). Here, $V=L\times L$ is the number of momenta in rBZ$_4$. We already find qualitative agreement between  4-OHK and the state-of-the-art DQMC simulations on the Hubbard model. This is expected given the rapid convergence of OHK to the Hubbard model. In particular, we observe the softening of the Mott gap as $N$ increases in both the density and compressibility. However, there are quantitative differences, for example, 4-OHK overestimates the magnitude of the Mott gap.

In Fig.~\ref{fig:density_compress}(e,f), we present the benchmark for double occupancy $\mathcal{D}$, and its derivative $\partial \mathcal{D}/ \partial \langle \hat{n} \rangle$ as a function of density $\langle \hat{n} \rangle$.  Double occupancy in the OHK model is defined as
\begin{align}
    \mathcal{D} = \frac{1}{2 n V} \sum_{\substack{\mathbf{k}\in{\rm rBZ}_4,\\ \alpha, \sigma \neq \sigma' }} \langle n_{\alpha \mathbf{k} \sigma} n_{\alpha \mathbf{k} \sigma'} \rangle .
\end{align}
This definition differs from that used in the SU($N$) Hubbard model.      
However, in both cases, $\mathcal{D}$ is proportional to the average interaction energy and the first derivative of the free energy with respect to $U$, thereby preserving this physical correspondence. In \figdisp{fig:density_compress}(e,f), the solid lines are the OHK results, while the markers denote DQMC-Hubbard results. The double occupancy from 4-OHK and DQMC-Hubbard agrees quantitatively. Moreover, 4-OHK shows consistent results at higher density $\langle \hat{n} \rangle>1$, whereas DQMC data becomes oscillatory due to the severe fermionic sign problem. As previously reported for SU($N$) Hubbard~\cite{ibarra2021}, the sharp jump in $\partial \mathcal{D} / \partial \langle \hat{n} \rangle$ serves as a strong signature of the Mott insulating state persisting even in the high-temperature regime. 

From this benchmarking, we find that the 4-OHK model successfully captures the high-temperature $N$-dependent physics in the SU($N$) Hubbard model. Aside from the overestimate of the Mott gap, we find quantitative agreement between 4-OHK and DQMC-Hubbard results for the density $\langle \hat{n}\rangle$-dependence of physical quantities, such as $\mathcal{D}$. Additionally, the simplicity of 4-OHK makes it advantageous for exploring higher densities and lower temperatures (next section), where DQMC simulations are limited by the fermionic sign problem. These features establish OHK as a powerful tool for studying SU($N$) Hubbard physics beyond the reach of conventional numerical methods.

{\bf Low temperature.} 
In \figdisp{fig:mott_gap}(a), we calculate the $\langle \hat{n} \rangle$ vs $\mu$ relation around $\langle \hat{n}\rangle=1$ at $U=8t$ and inverse temperature $\beta=200 t^{-1}$ for the 4-OHK model with varying $N=2,3,4$. As expected from the high-temperature results in \figdisp{fig:density_compress}(a,b), we observe a sharp characteristic plateau, signaling the presence of a Mott charge gap. However, a surprising feature emerges: the Mott gap does not shrink monotonically with increasing $N$ despite appearing to do so at high temperatures in both 4-OHK (\figdisp{fig:density_compress}(a)) or DQMC simulations of the Hubbard model (\figdisp{fig:density_compress}(b,f)). Fig.~\ref{fig:mott_gap}(a) reveals the ordering $\Delta(N=3)>\Delta(N=4)>\Delta(N=2)$
for $U=8t$, deviating from the high-temperature behavior. This discrepancy arises because, at high temperatures, the number of excited states increases monotonically with $N$, leading to more pronounced thermal softening of the gap for larger $N$. By leveraging the 4-OHK model, we can explore both low- and high-temperature regimes, allowing us to disentangle intrinsic ground-state properties from thermal effects associated with excited states.

Further, we calculate the charge gap at $\langle \hat{n}\rangle=1$ as a function of $U$ for various $N$ in the 4- and 8 (4$\times$2)-OHK models at zero temperature, as shown in \figdisp{fig:mott_gap}(b) and (c), respectively. In the SU(2) case, a charge gap opens for any $U>0$ due to Fermi surface nesting at exactly $\langle \hat{n}_\sigma\rangle=1/2$ with ${\bf q} = (\pi,\pi)$~\cite{feng2023,ibarra2023, mai2024}. For $N>2$, such a nesting is absent at $\langle \hat{n}\rangle=1$, where $\langle \hat{n}_\sigma\rangle=1/N<1/2$. Consequently, a finite critical $U_c$ is required to open the gap, as indicated in \figdisp{fig:mott_gap}(b). Moreover, we find that $U_c(N=3)>U_c(N=4)$, likely due to the particular Fermi surface geometry at $U=0$. This trend, $U_c(N=3)>U_c(N=4)$ is also observed in the 8-OHK (\figdisp{fig:mott_gap}) with only quantitative modification. Once the gaps are open for both SU($3$) and SU($4$) cases, we observe $\Delta(N=3)>\Delta(N=4)$, consistent with \figdisp{fig:mott_gap}(a). Extending our calculation to larger $N$, we find in \figdisp{fig:mott_gap}(b) the results collapse for $N>n$. This can be understood from the structure of the OHK model, which consists of independent $n$-site Hubbard clusters at each $\mathbf{k}$ in the rBZ$_n$. When a charge gap opens at $\langle \hat{n}\rangle=1$, these $n$-site clusters are filled with exactly $n$ particles each. When $N>n$, these particles occupy different SU($N$) flavors to minimize the kinetic energy. Further increasing $N$ does change this configuration, meaning that $n$-OHK would not be able to capture any additional $N$-dependent physics beyond $N=n+1$. For this reason, for the band HK and 2-OHK, the $\langle \hat{n}\rangle=1$ results overlap for all SU($N>1$) and SU($N>2$), respectively (see the appendix). It is important to note that this phenomenon is specific to the filling $\langle \hat{n}\rangle=1$. As we will discuss in the next section, near half-filling $\langle \hat{n}\rangle=N/2$ when the charge gap opens, each Hubbard cluster in OHK contains $Nn/2$ particles. In this case, $n$-OHK retains the non-trivial SU($N$) physics for any $N$ as long as $n \geq 2$.

\begin{figure}
       \begin{overpic}[width=0.8\columnwidth, keepaspectratio]{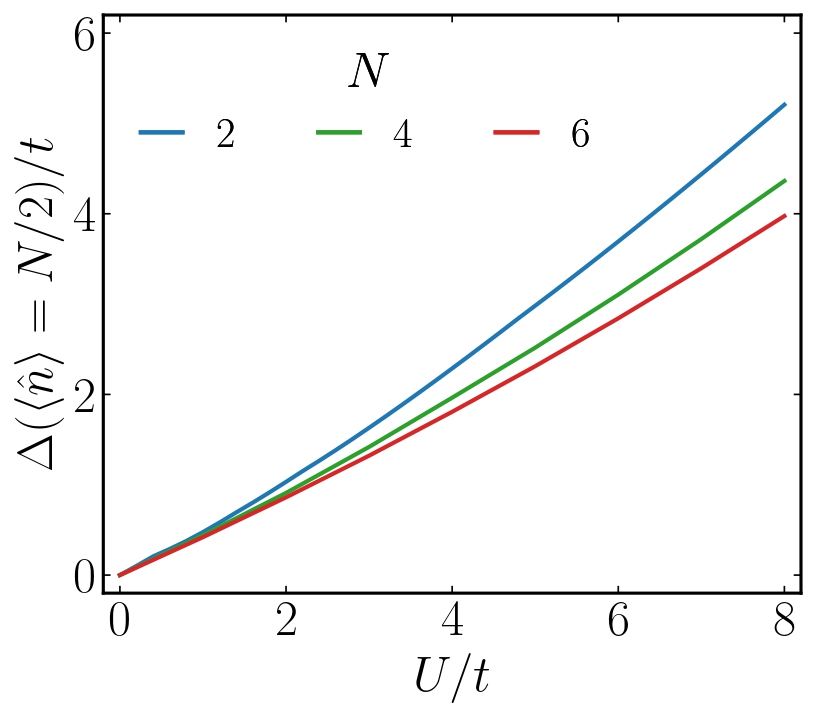}
  \end{overpic}
	\caption{Charge gap at half-filling ($\langle \hat{n} \rangle=N/2$) as a function of interaction strength $U$ for SU($N$) 4-OHK model.}
	\label{fig:hallfilled_mottgap}
\end{figure}

\subsection{Mott insulator at \texorpdfstring{$\langle \hat{n} \rangle = N/2$}{}}
As mentioned above, another generalization of the half-filled SU($2$) Mott insulator is the SU($N$) Mott insulator at half-filling $\langle \hat{n} \rangle = N/2$ for even $N$ (no half-filled charge gap exists for odd $N$). At this filling, the non-interacting Fermi surface for each flavor remains the same as in the SU($2$) case, thereby preserving the effect of Fermi surface nesting. Hence, we expect the charge gap opens for any finite $U>0$. This expectation is confirmed in \figdisp{fig:hallfilled_mottgap} where we present the half-filled charge gap as a function of $U$ for the SU(N) 4-OHK model. For a fixed interaction strength, the charge gap decreases monotonically with increasing $N$. As the number of flavors increases, more electrons participate in scattering processes, making it harder to enhance the single-particle charge gap with increasing $U$.

\begin{figure}
\centering
\includegraphics[width=0.8\columnwidth]{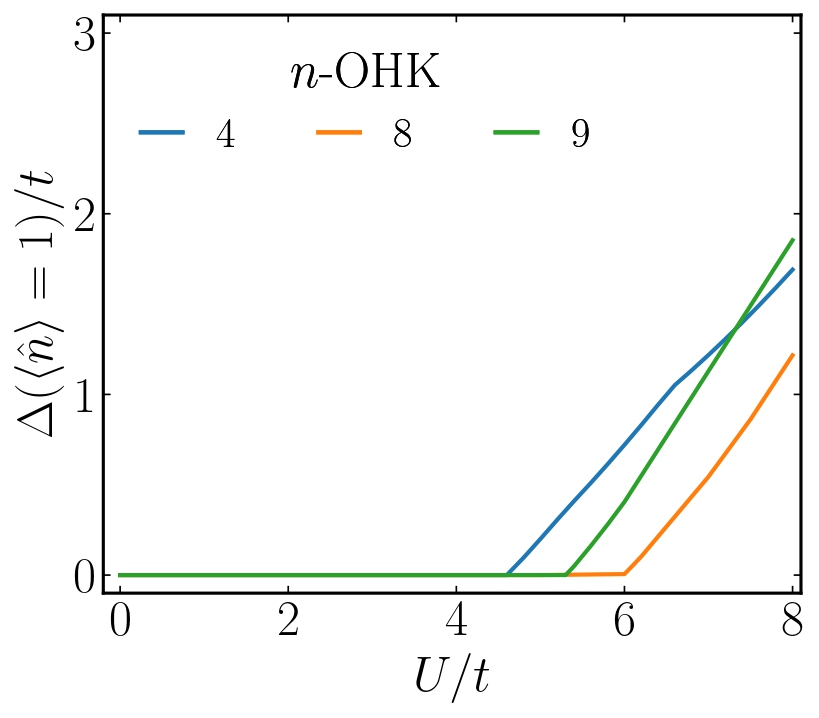}
	\caption{Mott gap at filling $\langle \hat{n} \rangle = 1$ for SU($3$) $n$-OHK as a function of interaction strength $U$ for varying $n$. For the 8-OHK model, we use a $4\times 2$ unit cell.}
	\label{fig:mottgap_N=3}
\end{figure}

\section{\label{sec:mag}SU(3) antiferromagnetism}

\begin{figure*}
\centering
\begin{overpic}[width=0.9\textwidth, keepaspectratio]{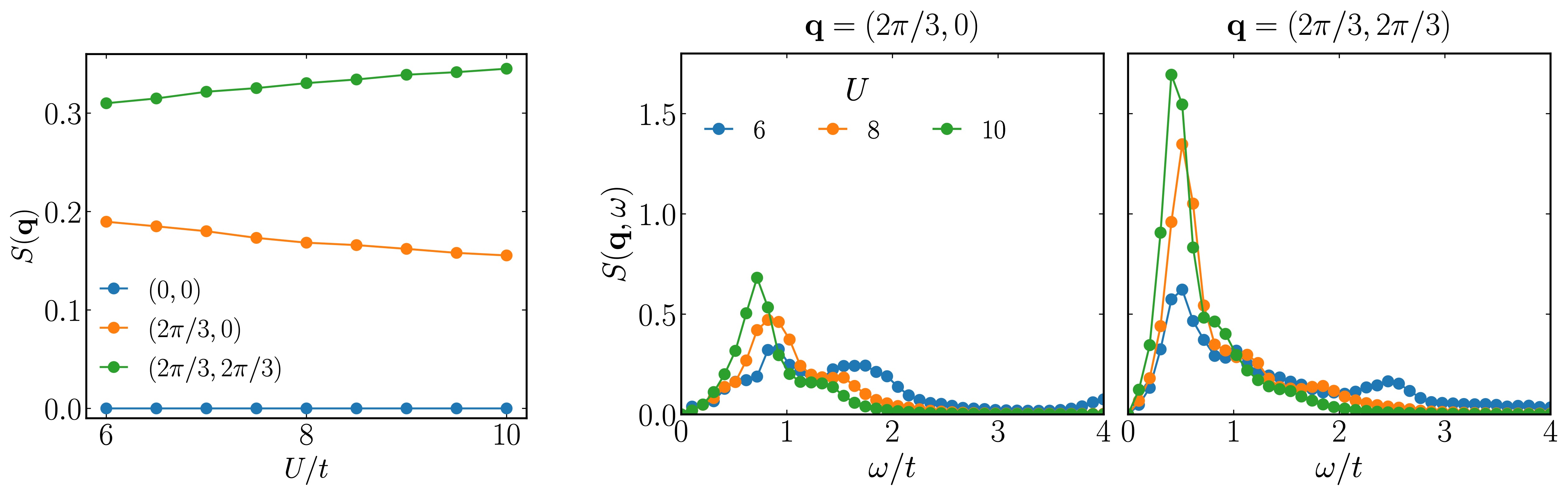}
    \put(2, 29){(a)}
    \put(40, 29){(b)}
  \end{overpic}
	\caption{Panel (a) shows the equal-time structure factor for SU($3$) nine-orbital HK as a function of the interaction strength $U$ at momenta $\mathbf{q} = (0,0)$, $(2\pi/3,0)$ and $(2\pi/3,2\pi/3)$. 
    Panel (b) shows the dynamical structure factor for  SU($3$) nine-orbital HK as a function of frequency $\omega$ at momenta ${\bf q} = (2\pi/3, 0)$ and $(2\pi/3, 2\pi/3)$ for $U=6,8$ and $10$. All calculations are carried out in the $\beta\rightarrow \infty$ limit and a broadening factor $\eta=0.1$ is used for dynamical calculations.}
	\label{fig:struc_comb}
\end{figure*}

Apart from charge properties, we are also interested in generalized magnetic correlations in the SU($N$) models. As previously studied~\cite{feng2023,ibarra2023}, we consider the equal-time structure factor for the same-spin or same-flavor species defined as
\begin{align}
    S(\mathbf{q})
    &= \frac{1}{n N V} \sum_{i j \sigma} \langle (n_{{\bf r}_i\sigma} - \langle n_{{\bf r}_i\sigma} \rangle ) ( n_{{\bf r}_j\sigma} - \langle n_{{\bf r}_j\sigma} \rangle )\rangle e^{i{\bf q} \cdot ({\bf r}_i - {\bf r}_j)}, 
\end{align}
where the transfer momentum $\bf q$ lies in the original full BZ. Constructing an OHK model based on a specific cluster scheme already selects a corresponding set of scattering momenta ${\bf q}_s$ (also within the original BZ) in the interaction term~\cite{mai2024}. We focus on ${\bf q}={\bf q}_s$ since only the scattering of these momenta are considered. Then $S({\bf q})$ adopts a simple form (see the appendix for a detailed derivation)
\begin{align}
    S({\bf q})
    = \frac{1}{n N V} \sum_{\substack{\mathbf{k}\in{\rm rBZ}_n, \\ \alpha\alpha',\sigma }}  &\big( \langle n_{\alpha {\bf k}  \sigma} n_{\alpha' {\bf k}  \sigma} \rangle \nonumber\\
    &- \langle  n_{\alpha {\bf k}  \sigma} \rangle \langle n_{\alpha' {\bf k}  \sigma} \rangle\big) e^{i {\bf q}\cdot ({\bf r}_\alpha-{\bf r}_{\alpha'})}, 
\end{align}
where ${\bf r}_\alpha$ denotes the position of the $\alpha$ orbital within the unit cell. 

Here we calculate zero-temperature $S(\mathbf{q})$ of the SU($3$) 9-OHK model ($3\times 3$ unit cell) at filling $\langle \hat{n} \rangle = 1$. The choice of $3\times3$ clustering is to include the scattering of momentum $(2\pi/3,2\pi/3)$, which has been shown to be the leading magnetic correlations in the SU(3) Hubbard model at a large $U$~\cite{feng2023,ibarra2023}. First, we determine the critical interaction strength for the metal-insulator transition in the 9-OHK model. In Fig~\ref{fig:mottgap_N=3}, we fix $N=3$ and plot the $\langle \hat{n} \rangle = 1$ Mott gap as a function of $U$ for varying orbital number $n=4,8,9$. We observed that the predicted $U_c$ is reasonably stable for the larger orbital numbers. For the 9-OHK, we find $U_c \simeq 5.4t$, which is in line with the Hubbard model results obtained using DQMC ($U_c \simeq 6t$)~\cite{ibarra2023} and constrained-path auxiliary-field QMC ($U_c \simeq 5.5t$)~\cite{feng2023}.

The $S(\mathbf{q})$ is shown in Fig.~\ref{fig:struc_comb}(a) as a function of $U>U_c$ at special momenta $\mathbf{q} = (0,0)$, $(2\pi/3, 0)$ and $(2\pi/3, 2\pi/3)$. 
We observe that the structure factor at $\mathbf{q} = (0,0)$ vanishes. For ${\bf q} =(2\pi/3,0)$, the structure factor is suppressed as the interaction strength increases. In contrast, the structure factor at ${\bf q} = (2\pi/3 , 2\pi/3)$ is enhanced. This result indicates that the OHK interaction favors $(2\pi/3 , 2\pi/3)$ antiferromagnetic (AFM) correlation, consistent with previous studies~\cite{feng2023}.

Another advantage of the OHK treatment over DQMC is the ease of obtaining dynamical correlations (the local spectral function of OHK is shown in the appendix and compared to dynamical mean-field theory~\cite{lee2018}). Here we consider the dynamical same-spin structure factor
\begin{align}
    S({\bf q},\omega) = -&\text{Im}\Bigg[\frac{1}{n N V} \sum_{\substack{\mathbf{k}\in{\rm rBZ}_n, \\ \alpha\alpha',\sigma }} \Big( \langle n_{\alpha {\bf k} \sigma} \frac{1}{\omega + i\eta + E_0 - H} n_{\alpha' {\bf k} \sigma} \rangle \nonumber\\
    &+\langle n_{\alpha' {\bf k} \sigma} \frac{1}{\omega - i\eta - E_0 + H} n_{\alpha' {\bf k} \sigma} \rangle \Big) e^{i {\bf q} \cdot ({\bf r}_\alpha - {\bf r}_{\alpha'})}\Bigg] ,
\end{align}
at zero temperature. This quantity is computed accurately using the Lanczos method with the ground state obtained from exact diagonalization. The dynamical structure factor as a function $\omega$ at momenta ${\bf q} = (2\pi/3, 0)$ and $(2\pi/3, 2\pi/3)$ is shown in Fig.~\ref{fig:struc_comb}(b) and (c), respectively, for varying interaction strengths $U=6,8,10$. With increasing $U$, we observe the clear development of a low-energy pole ($\omega_{\rm peak}(2\pi/3, 2\pi/3)\approx 0.4t$ for $U=10t$) for $S({\bf q} = (2\pi/3, 2\pi/3),\omega)$. In contrast, at ${\bf q} = (2\pi/3, 0)$, the correlator shows a relatively weaker peak centered at a slightly higher frequency ($\omega_{\rm peak}(0, 2\pi/3)\approx 0.7t$ for $U=10t$). This suggests the system has a propensity towards AFM magnetic ordering at wavevector $(2\pi/3, 2\pi/3)$ with sub-leading fluctuations at ${\bf q} = (2\pi/3, 0)$. This result agrees with the equal-time structure factor calculation.

The development of $(2\pi/3, 2\pi/3)$ AFM correlations at large $U$ under filling $\langle \hat{n} \rangle = 1$ is consistent with results from SU($3$) Hubbard~\cite{feng2023,ibarra2023} and Heisenberg models~\cite{toth2010,Bauer2012, romen2020}. The OHK model exhibiting the same magnetic behavior in the Mott insulating phase as the Hubbard model is another benchmark to support the validity of our approach.

The emergence of the low-energy pole at momenta $\mathbf{q}=( 2\pi/3, 2\pi/3)$ coincides with the critical interaction strength $U_c \simeq 5.4t$ which opens the Mott gap (see Fig.~\ref{fig:mottgap_N=3}). This suggests that the magnetic ordering and metal-insulator transition are correlated: at small interaction strengths $U < U_c$, the system is in a non-magnetic metallic state, and, as we increase the interaction strength $U > U_c$, the system undergoes a transition to a Mott insulator with leading magnetic correlation with $\mathbf{q}=(\pm2\pi/3,\pm2\pi/3)$ at low energy and sub-leading correlation with $\mathbf{q}=(\pm2\pi/3,0) \text{and} (0,\pm2\pi/3)$ at slightly higher energy (here we included the symmetric momenta). 

\section{\label{sec:discussion}Discussion}

We have shown how SU($N$) physics can be teased out of the Hubbard model using the OHK methodology.  For a modest number of orbitals, we can reproduce the DQMC results quantitatively for double occupancy and qualitatively for filling and compressibility. Our results demonstrate that the OHK scheme is a reliable method for studying the physics of strong correlations without the full computational machinery of DQMC. A major advantage of OHK is that low-temperature physics and dynamics are readily available without analytical continuation. Consequently, it is now a reasonable goal to obtain transport properties of the SU($N$) model as a function of temperature. A project along these lines is now being pursued.

 {\bf Data availability statement:} The data that support the findings of this article are openly available~\cite{hackner_2025_15476864}.

\begin{acknowledgments}

We thank Eduardo Ibarra-García-Padilla {\it et al.} for sharing their DQMC data plotted in Fig.~\ref{fig:density_compress}(b,d,e,f), and Kaden R. A. Hazzard and Gabriele la Nave
for suggesting we work on this problem and subsequent useful discussions. P.W.P. acknowledges NSF DMR-2111379 for partial funding. P.M. was supported by the Gordon and Betty Moore Foundation’s EPiQS Initiative through grant GBMF 8691.

\end{acknowledgments}

\appendix

\section{\label{sec:bandDerivation}Analytic results for SU(\texorpdfstring{$N$}{}) band HK}
Here we demonstrate that the exact solvability of the band HK model is inherited by the generalized SU($N$) model. The SU($N$) band HK model is simply the single orbital limit of the general model given in Eq.~\ref{eq:SU(N)OHK}
\begin{align}
    H_\text{HK}^N = \sum_\sigma \sum_{\bf k} \xi_{\bf k} n_{{\bf k}\sigma} + \frac{U}{2}\sum_{\sigma\neq\sigma^\prime} \sum_{\bf k} n_{{\bf k}\sigma} n_{{\bf k}\sigma^\prime}.
\end{align}
The Hamiltonian is diagonalized by the Fock space basis $|n_{{\bf k}1},...,n_{{\bf k}N}\rangle$. At each momentum ${\bf k}$, the corresponding energies are given by
\begin{align}
    E^{n_{\bf k}}_{\bf k} = n_{\bf k} \xi_{\bf k} + \frac{n_{\bf k}(n_{\bf k}-1)}{2} U ,
\end{align}
where $n_{\bf k}$ denotes the number of filled states at momentum ${\bf k}$. Each energy has a degeneracy of $\begin{pmatrix}
    N\\
    n_{\bf k}
\end{pmatrix}$, and so the partition function can be written
\begin{align}
    Z_{\bf k} = \sum_{n_{\bf k}=0}^N \begin{pmatrix}
        N \\
        n_{\bf k}
    \end{pmatrix} e^{-\beta E^{n_{\bf k}}_{\bf k}}.
\end{align}
The thermally-weighted filling and double occupancy at momentum $\mathbf{k}$ are given by
\begin{align}
    \langle \hat{n}_{\bf k}  \rangle = \frac{1}{Z_{\bf k}} \sum_{n_{\bf k}=0}^N \begin{pmatrix}
        N \\
        n_{\bf k}
    \end{pmatrix} n_{\bf k} e^{-\beta E^{n_{\bf k}}_{\bf k}},
\end{align}
and
\begin{align}
    \mathcal{D}_{\bf k} = \frac{1}{Z_{\bf k}} \sum_{n_{\bf k}=0}^N \begin{pmatrix}
        N\\
        n_{\bf k}
    \end{pmatrix} \frac{n_{\bf k}(n_{\bf k}-1)}{2} e^{-\beta E^{n_{\bf k}}_{\bf k}},
\end{align}
respectively.

The Green function of the system is exactly solvable for all $N$. Noting that $c_{{\bf k}\sigma}(\tau) = e^{-(\xi_{\bf k} + U \sum_{\sigma^\prime\neq\sigma} n_{{\bf k}\sigma^\prime} )\tau} c_{{\bf k}\sigma}$, we have that
\begin{align}
    G_{{\bf k}\sigma}(\tau) &= -\langle c_{{\bf k}\sigma}(\tau) c_{{\bf k}\sigma}^\dagger \rangle \nonumber\\
    &= -\frac{1}{Z_{\bf k}} \text{Tr} \left[ e^{-\beta H} e^{-(\xi_{\bf k} + U \sum_{\sigma^\prime\neq\sigma} n_{{\bf k}\sigma^\prime} )\tau}  (1-n_{{\bf k}\sigma}) \right] \nonumber\\
    &= -\frac{1}{Z_{\bf k}} \sum_{n_{\bf k}=0}^{N-1} \begin{pmatrix}
        N-1\\
        n_{\bf k}
    \end{pmatrix} e^{-\beta E^{n_{\bf k}}_{\bf k}} e^{-(\xi_{\bf k} + U n_{\bf k})\tau} ,
\end{align}
and the corresponding Matsubara Green function 
\begin{align}
     G_{{\bf k}\sigma} (i \omega_m) = \frac{1}{Z_{\bf k}} \sum_{n_{\bf k}=0}^{N-1} \begin{pmatrix}
        N - 1 \\
        n_{\bf k}
    \end{pmatrix} \frac{e^{-\beta E^{n_{\bf k}+1}_{\bf k}}+e^{-\beta E^{n_{\bf k}}_{\bf k}}}{i\omega_m - (\xi_{\bf k} + U n_{\bf k})}. 
\end{align}

\section{\label{sec:observables} Computing observables}
For a generic operator $\hat{O}$, it's thermally-weighted expectation value is given by
\begin{align}
    \langle \hat{O} \rangle = \frac{1}{Z} \text{Tr} [ \hat{\rho} \hat{O}], \quad \hat{\rho} = e^{-\beta \hat{H}},\quad Z = \text{Tr} [\hat{\rho}] .
\end{align}
The OHK Hamiltonian decomposes into independent clusters at each $\bf k$ in the reduced Brillouin zone (rBZ$_n$), and thus takes the form $\hat{H} = \sum_\mathbf{k} \hat{H}_\mathbf{k}$. This allows the partition function to factorize as
\begin{align}
Z = \prod_{{\bf k}\in \text{rBZ}_n} Z_{\bf k}=\prod_{{\bf k}\in \text{rBZ}_n} \text{Tr} [\hat{\rho}_{\bf k}], \quad \hat{\rho}_{\bf k}=e^{-\beta \hat{H}_{\bf k}}.
\end{align}
The thermal expectation values can be evaluated separately at each $\bf k$ and then summed over $\bf k$
\begin{align}
\langle \hat{O} \rangle = \sum_{{\bf k}\in \text{rBZ}_n} \langle \hat{O} \rangle_{\bf k}, \quad \langle \hat{O} \rangle_{\bf k}=\frac{1}{Z_{\bf k}} \text{Tr} [ \hat{\rho}_{\bf k} \hat{O}].
\end{align}

\section{\label{sec:2OHK} Numerical results for 2-OHK}
\begin{figure*}
  \begin{overpic}[width=\textwidth, keepaspectratio]{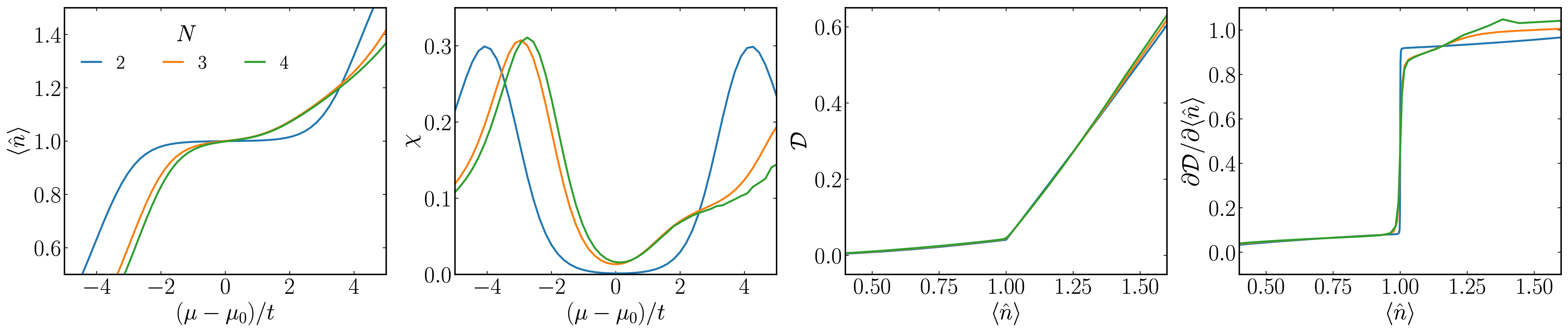}
    \put(5,  19){(a)}
    \put(30,  19){(b)}
    \put(55,  19){(c)}
    \put(80,  19){(d)}
  \end{overpic}
    \caption{High temperature thermodynamic calculations for SU($N$) 2-OHK near the $\langle \hat{n} \rangle=1$ Mott gap for varying $N=2,3,4$. Panels (a) and (b) show $\langle \hat{n} \rangle$ and $\chi$ versus chemical potential $\mu$, respectively. Panels (c) and (d) show the $\mathcal{D}$ and $\partial\mathcal{D}/\partial\langle\hat{n}\rangle$ versus $\langle\hat{n}\rangle$, respectively. For (a) and (b) we calculate $\mu_0$ such that $\langle \hat{n} \rangle |_{\mu_0} = 1$, and for all calculations we fix $U=8t$ and $\beta=2t^{-1}$. Simulated on a $L\times L$ square lattice with $L=20$.}
	\label{fig:finite_temp}
\end{figure*}

\begin{figure*}
  \begin{overpic}[width=0.8\textwidth, keepaspectratio]{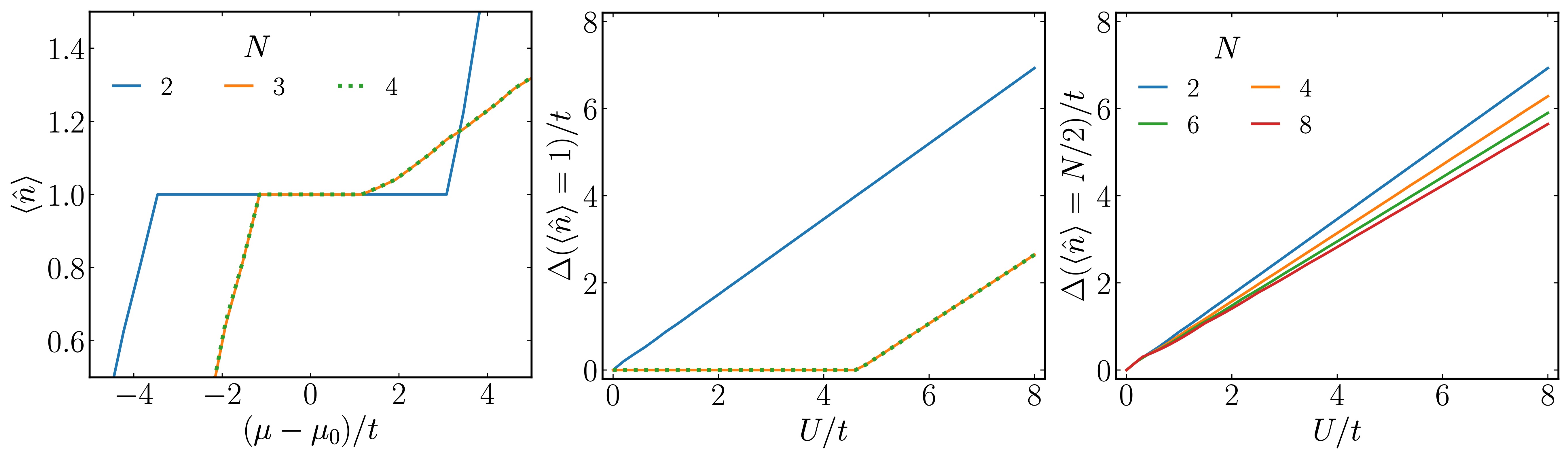}
    \put(7,  26){(a)}
    \put(40,  26){(b)}
    \put(73,  26){(c)}
  \end{overpic}
    \caption{Low temperature calculations for SU($N$) 2-OHK. Panel (a) shows density $\langle \hat{n} \rangle$ versus $\mu$ near the $\langle \hat{n}\rangle=1$ Mott gap with $U=8t$ and $\beta=200t^{-1}$. Panels (b) and (c) show the $\langle \hat{n}\rangle=1$ and $\langle \hat{n}\rangle=N/2$ Mott gaps, respectively, as a function of $U$. Note that (a) and (b) share a legend. Simulated on a $L\times L$ square lattice with $L=20$.}
	\label{fig:zero_temp}
\end{figure*}

For completeness, we provide the 2-OHK results corresponding to quantities calculated in Figs.~1-4 for 4-OHK in the main text. In Fig.~\ref{fig:finite_temp}, we show the density $\langle\hat{n}\rangle$, compressibility $\chi$, double occupancy $\mathcal{D}$ and $\partial\mathcal{D} / \partial \langle \hat{n} \rangle$ near the $\langle\hat{n}\rangle =1$ Mott gap with $U=8t$ at high temperature $\beta=2t^{-1}$. In Fig.~\ref{fig:zero_temp}(a), we show $\langle\hat{n}\rangle$ versus $\mu$ near the $\langle\hat{n}\rangle =1$ Mott gap for $U=8t$ at low temperature $\beta=200t^{-1}$. In Fig.~\ref{fig:zero_temp}(a) and (b), we show the $\langle \hat{n}\rangle=1$ and $\langle \hat{n}\rangle=N/2$ Mott gaps, respectively, as a function of $U$.

\section{Derivation of same-spin structure factor in the orbital basis}
We consider the equal-time same-spin structure factor defined by
\begin{widetext}
    \begin{align}
   S(\mathbf{q})
    &= \frac{1}{n N V} \sum_{i j \sigma} \langle (n_{{\bf r}_i\sigma} - \langle n_{{\bf r}_i\sigma} \rangle ) ( n_{{\bf r}_j\sigma} - \langle n_{{\bf r}_j\sigma} \rangle ) \rangle e^{i{\bf q} \cdot ({\bf r}_i - {\bf r}_j)}  \nonumber\\
    &= \frac{1}{n N V} \sum_{ij\sigma} \sum_{\alpha\alpha'} \left( \langle n_{\alpha {\bf R}_i\sigma} n_{\alpha'{\bf R}_j\sigma} \rangle - \langle n_{\alpha{\bf R}_i\sigma} \rangle \langle n_{\alpha'{\bf R}_j\sigma} \rangle \right) e^{i {\bf q} \cdot \left( ({\bf R}_i - {\bf R}_j) + ({\bf r}_\alpha-{\bf r}_{\alpha'})\right)} ,
\end{align}
where $n,\ N$ and $V$ are the number of orbitals, spin flavours and unit cells, respectively. In the last line, we have changed to the orbital basis where ${\bf R}_i$ and ${\bf r}_\alpha$ denote the position of the unit cell and the position of the orbital within the unit cell, respectively. Note that the scattering momentum $\bf q$ can lie anywhere in the BZ. 
We now define the Fourier transform of the real-space density operator
\begin{align}
    n_{\alpha {\bf R}_i\sigma} &= \frac{1}{V} \sum_{\bf q \in \text{rBZ}_n} \rho_{\alpha{\bf q}\sigma} e^{-i {\bf q} \cdot ({\bf R}_i+{\bf r}_\alpha)}, \\
    \rho_{\alpha{\bf q}\sigma} &= \sum_{i} n_{\alpha {\bf R}_i\sigma} e^{i\bf q \cdot (R_i + r_\alpha)} = \sum_{\bf k \in \text{rBZ}_n} c^\dagger_{\alpha {\bf k+q} \sigma} c_{\alpha {\bf k} \sigma}.
\end{align}
In the following, all momentum summations are restricted to the rBZ$_n$. Replacing the real-space operators, we have
\begin{align}
    S({\bf q}) &= \frac{1}{n N V^3} \sum_{ij\sigma}\sum_{\alpha\alpha'} \sum_{{\bf k k^\prime}} \left( \langle \rho_{\alpha {\bf k}^\prime\sigma} \rho_{\alpha' {\bf k} \sigma} \rangle - \langle \rho_{\alpha {\bf k}^\prime\sigma} \rangle \langle \rho_{\alpha' {\bf k} \sigma} \rangle \right) e^{-i ({\bf k}^\prime\cdot ({\bf R}_i+{\bf r}_\alpha) + {\bf k}\cdot ({\bf R}_j + {\bf r}_{\alpha'}))} e^{i{\bf q}\cdot \left( ({\bf R}_i - {\bf R}_j) + ({\bf r}_\alpha-{\bf r}_{\alpha'})\right)} \nonumber\\
    &= \frac{1}{n N V} \sum_{\sigma}\sum_{\alpha\alpha'} \sum_{\bf k k^\prime} \left( \langle \rho_{\alpha {\bf k}^\prime\sigma} \rho_{\alpha' {\bf k} \sigma} \rangle - \langle \rho_{\alpha {\bf k}^\prime\sigma} \rangle \langle \rho_{\alpha' {\bf k} \sigma} \rangle \right) \delta^n_{{\bf k}^\prime, {\bf q}} \delta^n_{{\bf k}, -{\bf q}} e^{-i ({\bf k}^\prime\cdot {\bf r}_\alpha + {\bf k}\cdot{\bf r}_{\alpha'})} e^{i{\bf q}\cdot ({\bf r}_\alpha-{\bf r}_{\alpha'})}, \nonumber\\
\end{align}
where $\delta^n_{\bf k,q}$ represents a delta function mod reciprocal lattice vectors for a given $n$-orbital HK model. Let ${\bf q = q^\prime + q_s}$ for $\bf q^\prime$ in the first BZ and the corresponding reciprocal lattice vector $\bf q_s$. Then we have $\bf q - k^\prime = q_s$ and $\bf -k - q = -q_s$
\begin{align}
     S({\bf q}) &= \frac{1}{n N V} \sum_{\sigma}\sum_{\alpha\alpha'} \left( \langle \rho_{\alpha {\bf q}^\prime\sigma} \rho_{\alpha' -{\bf q}^\prime \sigma} \rangle - \langle \rho_{\alpha {\bf q}^\prime\sigma} \rangle \langle \rho_{\alpha' -{\bf q}^\prime \sigma} \rangle \right) e^{i{\bf q_s}\cdot ({\bf r}_\alpha-{\bf r}_{\alpha'})} .
\end{align}
If we pick special scattering momenta such that $\bf q = q_s$, then $\bf q^\prime =0$
\begin{align}
     S({\bf q}) &= \frac{1}{n N V} \sum_{\sigma}\sum_{\alpha\alpha'}\left( \langle \rho_{\alpha {\bf 0} \sigma} \rho_{\alpha' {\bf 0} \sigma} \rangle - \langle \rho_{\alpha {\bf 0} \sigma} \rangle \langle \rho_{\alpha' {\bf 0} \sigma} \rangle \right) e^{i{\bf q}\cdot ({\bf r}_\alpha-{\bf r}_{\alpha'})} \nonumber\\
     &= \frac{1}{n N V} \sum_{\sigma}\sum_{\alpha\alpha'}\sum_{\bf k k^\prime } \left( \langle n_{\alpha {\bf k}^\prime\sigma} n_{\alpha' {\bf k} \sigma} \rangle - \langle n_{\alpha {\bf k}^\prime\sigma} \rangle \langle n_{\alpha' {\bf k} \sigma} \rangle \right) e^{i \bf q \cdot (r_\alpha-r_{\alpha'})} \nonumber\\
     &= \frac{1}{n N V} \sum_{\sigma}\sum_{\alpha\alpha'}\sum_{\bf k} \left( \langle n_{\alpha {\bf k}\sigma} n_{\alpha' {\bf k} \sigma} \rangle - \langle n_{\alpha {\bf k}\sigma} \rangle \langle n_{\alpha' {\bf k} \sigma} \rangle \right) e^{i \bf q \cdot (r_\alpha-r_{\alpha'})},  \nonumber\\
\end{align}
where $n_{\alpha{\bf k}\sigma} = c^\dagger_{\alpha {\bf k} \sigma}c_{\alpha {\bf k} \sigma}$ and on the last line we have used that OHK factorizes in momentum space in the orbital basis.
The corresponding dynamical structure factor is given by
\begin{align}
    S({\bf q},\omega) &= -\text{Im} \left[ \frac{1}{n N V} \sum_{\sigma}\sum_{\alpha \alpha'}\sum_{\bf k} \left( \langle  n_{\alpha {\bf k}\sigma} \frac{1}{\omega + i\eta + E_0 - H}  n_{\alpha' {\bf k}\sigma} \rangle + \langle  n_{\alpha {\bf k}\sigma} \frac{1}{\omega - i\eta - E_0 + H}  n_{\alpha' {\bf k}\sigma} \rangle \right) e^{i \bf q\cdot (r_\alpha-r_{\alpha'})} \right],
\end{align}
where we assume zero temperature and a non-degenerate ground state.
\end{widetext}

\begin{figure*}[t!]
  \begin{overpic}[width=0.8\textwidth, keepaspectratio]{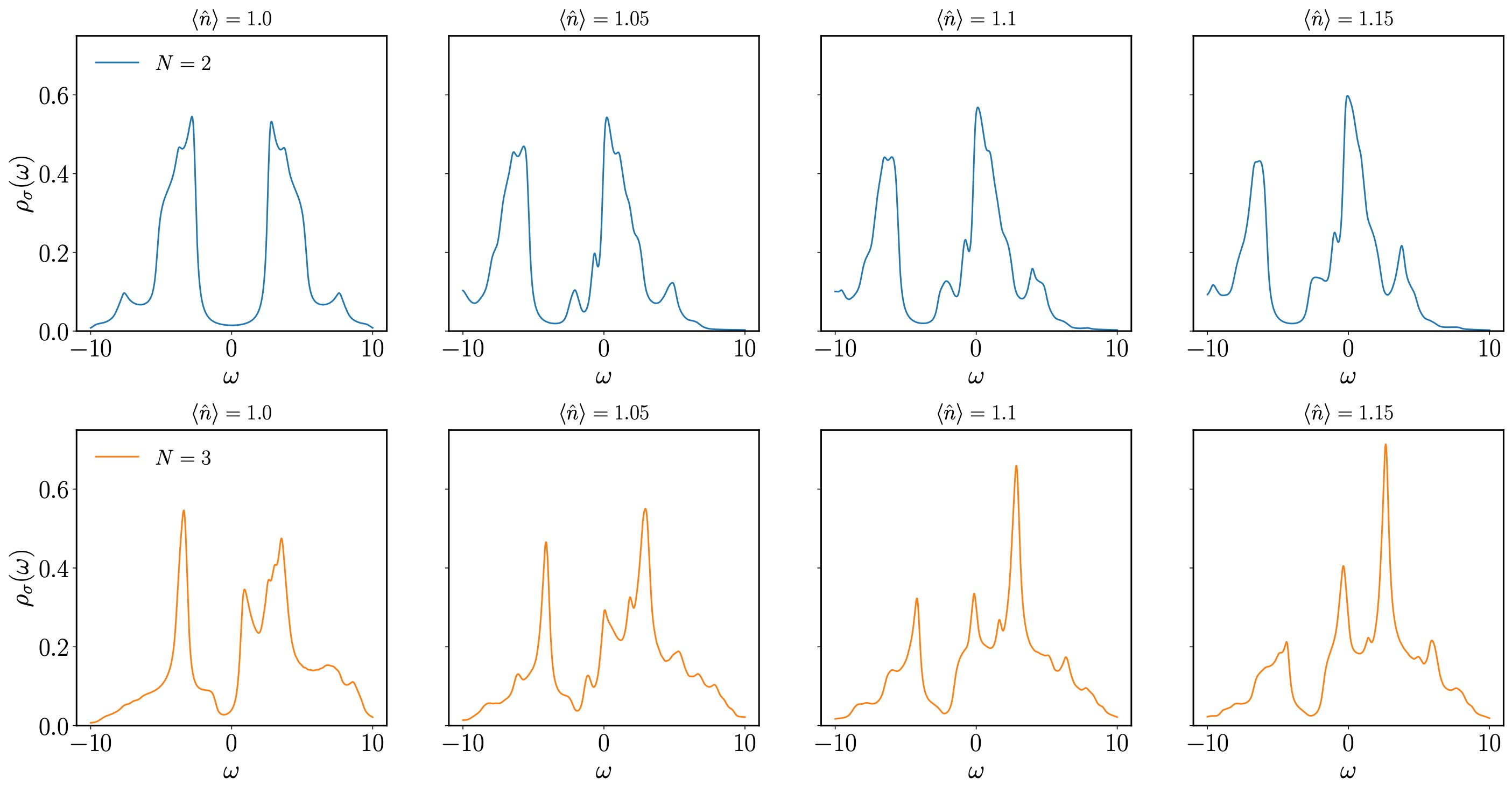}
    \put(2.5,  51){(a)}
    \put(27,  51){(b)}
    \put(51.5,  51){(c)}
    \put(76,  51){(d)}
    \put(2.5,  25){(e)}
    \put(27,  25){(f)}
    \put(51.5,  25){(g)}
    \put(76,  25){(h)}
  \end{overpic}
    \caption{ Panels (a-d) and (e-h) show the local spectral function versus $\omega$ for varying fillings for SU($N$) 4-OHK with $N=2$ and $N=3$, respectively. Calculations are carried out with $U=8t$, $\beta=200t^{-1}$ and broadening factor $\eta=0.2$. Simulated on a $L\times L$ square lattice with $L=40$.}
	\label{fig:spectral}
\end{figure*}

\section{\label{sec:spectral_plots} Local spectral function}
Here, we calculate the local spectral function for a single spin species, given by
\begin{widetext}
    \begin{align}
    \rho_\sigma(\omega) = -\text{Im} \left[ \frac{1}{n V}\sum_{\alpha,\bf k} \sum_{m n} \frac{\bra{n}c^\dagger_{\alpha\bf k \sigma} \ket{m} \bra{m} c_{\alpha\bf k \sigma} \ket{n} }{\omega - (E_{\bf k}^n - E^m_{\bf k}) + i\eta} \frac{e^{-\beta E_{\bf k}^n}+e^{-\beta E_{\bf k}^m}}{Z_{\bf k}} \right]
\end{align}
\end{widetext}
where $E_{\bf k}^n$ are the eigenvalues at momentum $\bf k$ and $Z_{\bf k} = \sum_n e^{-\beta E_{\bf k}^n}$. The $\rho_\sigma(\omega)$ at $\beta=200t^{-1}$ for SU($N=2,3$) are shown in Fig.~\ref{fig:spectral} as we particle dope the system away from one particle per site ($\langle n\rangle=1$). The emergence of a three-peak structure at small particle doping is in qualitative agreement with the dynamical mean-field theory (DMFT) results from Ref.~\cite{lee2018}. However, we notice a quantitative difference in the local spectral function between 4-OHK and DMFT, especially in the SU(3) case. In DMFT, the central peak at the chemical potential (zero-frequency) is the highest and sharpest feature. In contrast, in the 4-OHK model, while a central peak does emerge with increasing doping, the dominant spectral peak is located at a finite positive frequency. Since DMFT is only exact in the infinite $d$ limit, where momenta are decoupled, the 4-OHK model offers a valuable alternative and simple perspective to access the two-dimensional Hubbard physics with explicit momentum-mixing.

\bibliography{refs}

\end{document}